\documentclass[]{aastex631}

\newcommand{\unit}[1]{\left[{#1}\right]}{}
\newcommand{\gcms}{\text{g}/\text{cm}^3}{}
\newcommand{\cm}{{\text{cm}}}{}
\newcommand{\Ni}{${}^{56}$Ni}{}
\newcommand{\rhoint}{\rho_\text{interaction}}{}

\usepackage{amsmath}	

\begin{document}

\title{Thermonuclear supernova explosion criteria for direct and indirect collisions of CO white dwarfs: a study of the impact-parameter threshold for detonation}

\correspondingauthor{Hila Glanz}
\email{glanz@campus.technion.ac.il}

\author[0000-0002-6012-2136]{Hila Glanz}
\affiliation{Technion - Israel Institute of Technology \\
Haifa, 3200002, Israel}

\author[0000-0002-5004-199X]{Hagai B. Perets}
\affiliation{Technion - Israel Institute of Technology \\
Haifa, 3200002, Israel}

\author[0000-0003-3308-2420]{Ruediger Pakmor}
\affiliation{Max-Planck-Institut für Astrophysik\\ 
Karl-Schwarzschild-Str. 1, D-85748 \\
Garching, Germany}

\begin{abstract}
The physical collisions of two white dwarfs (WDs) (i.e. not slow mergers) have been shown to produce type-Ia-like supernovae (SNe) explosions. Most studies of WD collisions have focused on zero impact-parameter (direct) collisions, which can also be studied in 2D. However, the vast majority of WD collisions arising from any evolutionary channels suggested to date are expected to be indirect, i.e. have a non-negligible impact parameter upon collision. Here, we use one of the highest resolution 3D simulations to date (making use of the \texttt{AREPO} code) in order to explore both direct and indirect collisions and the conditions in which they give rise to a detonation and the production of a luminous SNe. Using our simulations, we find a detonation criterion that can provide the critical impact parameter for an explosion to occur, depending on the density profile of the colliding WDs, their composition, and their collision velocities.
We find that the initial velocity has a significant impact on the amount of \Ni{} production from the explosion. Furthermore, the production of the \Ni{} also depends on numerical modeling aspects.

\end{abstract}




\section{Introduction}
Type Ia supernovae are thought to be the result of thermonuclear explosions of carbon-oxygen (CO) white dwarfs (WDs) \citep{1960HoyleFowler}.
However, the exact mechanism in which such explosions occur and the exact progenitors are still debated.
A wide range of scenarios for the origins of type Ia SNe has been suggested and explored, and it is possible that more than one evolutionary channel can give rise to type Ia SNe (see reviews by \citealt{2014Maoz,2018Livio,2019Jha,2019Soker,2020Ruiter}). Many of these include the strong interactions between two WDs (double degenerate channels), either through a merger, or mass-transfer, or a direct collision. 

In particular, the direct collision between two WDs, i.e., a fast physical collision rather than a partial mass transfer and/or inspiral and merger, has been suggested to produce type Ia SNe. Such direct physical collisions are not easily produced in Nature. They can occur rarely through random encounters in dense stellar regions \citep{1993SigurdssonPhinneyGlobularClusterTripleInteractions, 2009RosswogWDCollision,2009RaskinWDCollisionSimulation}, and possibly somewhat more frequently in triple or quadruple systems \citep{2011ThompshonTripleMergers,2012KatzDongCollisionHighRate, 2012PeretsKratterTripleToCollisions, 2013KushnirKatzHeadOnWDCollisionsInTriple, 2018ToonenRateHeadOnWDCollisions, 2018HamersQuadruples, 2021MichaelyErezTypeIaInTheField}.
However, detailed population synthesis studies suggest that they are relatively rare and likely do not produce more than 1 percent of the type Ia SN rate, nor do they follow the delay time distribution for Ia-SNe inferred from observations \cite{2018ToonenRateHeadOnWDCollisions}. 
Nevertheless, these types of events are likely to occur in Nature and produce thermonuclear SNe. Their modeling can teach us both about physical processes occurring in thermonuclear SNe, and their exact contribution to the production of normal and peculiar SNe through this channel.

Previous works on Ia-SNe from WD collisions focused mainly on the head-on scenario of CO WDs \citep{1989Benz3DSimulationsWDCollisions,2009RosswogWDCollision, 2013KushnirKatzHeadOnWDCollisionsInTriple}, and also CO WD with Helium shells \citep{2016kushnirHeliumShells,2016PapishPeretsHeadOnCollision}. \cite{2014KushnirKatzSelfSimilarHeadOnSolution} developed a self-similar solution for the shock in the head-on collision scenario.
A few studies explored the change in the ${}^{56}$Ni production due to different impact parameters \citep{2009RaskinWDCollisionSimulation, 2010RaskinNi56FromWDCollisions, 2015MNRASDongKatzKushnirHighResHeadOnBiModal}, where \cite{2010LorenSPHWDCollisions} explored the case of collisions in low-mass WD binaries. These studies analyze the \Ni{} production of the different setups. However, a systematic investigation between the impact parameter and whether or not the system undergoes an explosion and quantifying this relation was not yet done (although \citealt{2010RaskinNi56FromWDCollisions} and \cite{2010LorenSPHWDCollisions} found some cases in their simulations that avoided explosions). 

Although the rate of Ia-SNe from head-on collisions might explain only a small fraction of the total rate of such events \citep{2018ToonenRateHeadOnWDCollisions}, collisions with non-zero impact parameters can result in a different element production, in particular a lower \Ni{} production, different light curve, and can even completely avoid the explosion that would have happened in the head-on scenario. 

In this paper, we simulate equal mass CO WD collisions in 3D with different impact parameters, in order to find the condition that is required for a full explosion to occur (leaving no remnant behind) and to study the differences in the productions of these collisions- the amount of \Ni, and remnants. For simplicity, and given the significant computational expense, we chose to focus here only on the case of collision of equal mass WDs. The important case of non-equal mass collisions is beyond the scope of the current study. We first examine the result of a direct (head-on) collision, and then increase the impact parameter according to the densities of the interacting surfaces, such that we find a condition that leads to only a small production of ${}^{28}$Si but not ${}^{56}$Ni and no explosion. With smaller impact parameters, we get an explosion, and with slightly larger, only a small perturbation, without any significant element production.

\section{Density threshold for ignition and the corresponding impact parameter}
\label{explosion criteria}
\begin{figure}
    \centering
    \includegraphics[width=0.7\linewidth,clip]{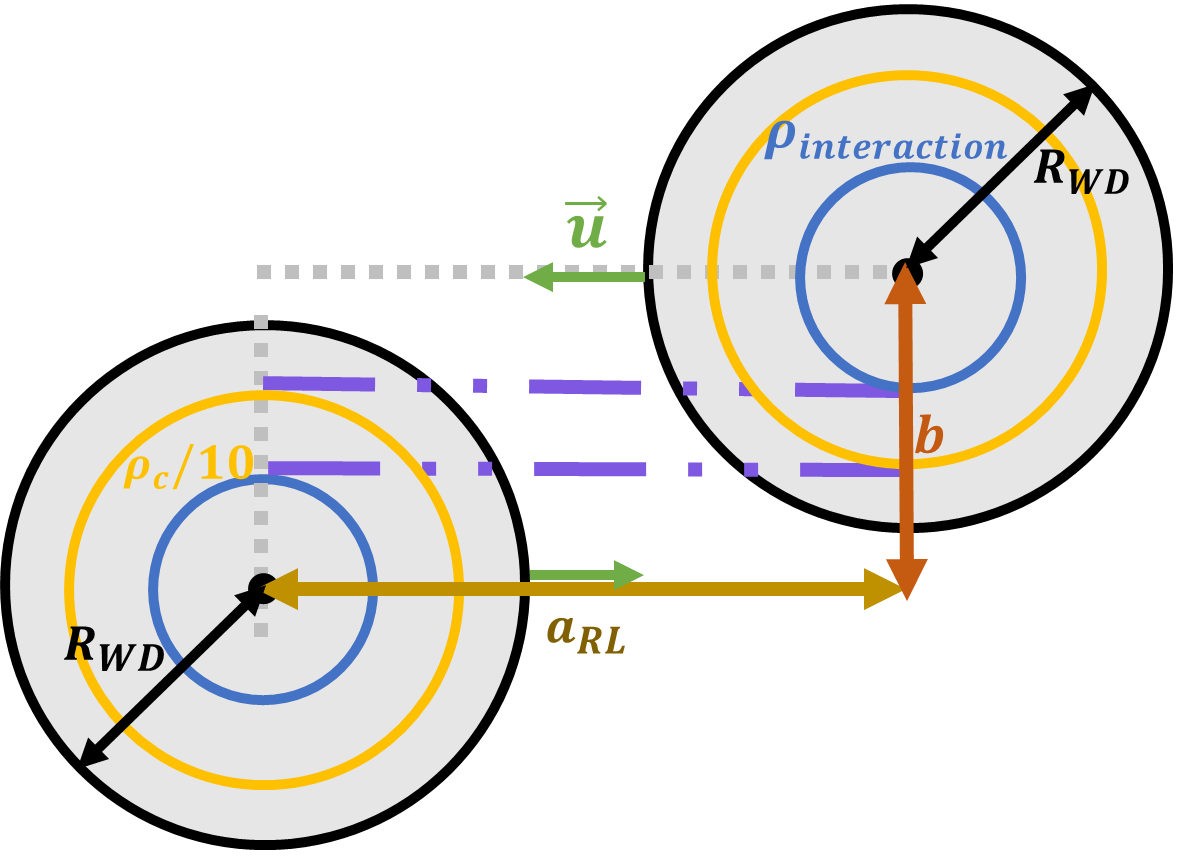}
    \caption{Initial configuration of the binary system, showing the definition of the interaction layer with $\rhoint$ which is the maximum density in the region of immediate impact, i.e.- the value of the deepest layer along the intersected area in the direction of motion. This is the area between the two purple lines, each connecting one WD with the edge of the colliding companion, where the density is high enough to be able to compress another layer with the critical density we calculate here. The values are the initial values at the beginning of the simulation at about the Roche-lobe size distance in the x-direction. We note that these values are changed due to gravitational focusing before the actual impact. Therefore, $\rhoint$ upon impact can be an order of magnitude larger compared to the initial configuration. In our calculations of the critical values of $\rhoint$ and the relative velocities for explosions, the values of interest are thus those found at the moment of impact, i.e., when the two WDs initially touch each other, and not their original values upon initialization.}
    \label{fig:sketch_initial_configuration}
\end{figure}

As two WDs collide (i.e. regions of the WD with high densities interact and rise to physical collision) with a relative velocity that is greater than their sound speed,  a supersonic shock wave occurs and propagates throughout the objects, increasing the densities, potentially igniting the carbon in one, or both WDs, leading to an explosion.
One can follow the Rankine-Hugoniot relations \citep{1870Rankine, hugoniot1887memoire,zeldovich1950theory,vonNeuman1942theory,1943Doring} to calculate the minimal initial (i.e. before any collision and shock happen) velocity and density in the collision region ($\rhoint$ in Fig. \ref{fig:sketch_initial_configuration}), such that the shock, resulting from the collision, can sufficiently compress the WD as to achieve the critical density $\rho_c$, and the temperature $T_c$ required for carbon ignition. As can be seen in Tab. \ref{tab:simulations_parameters}, the WDs change their direction of motion and impact parameters before the actual collision. The impact velocity is calculated according to the motion axis at the moment the WDs physically impact each other, and so is the impact parameter.

We define the density at the interacting layer, $\rhoint$, to be the maximum density in the volume directly interacting in the collision, i.e. between the two purple lines of Fig. \ref{fig:sketch_initial_configuration}- the lines connecting one WD surface with a high enough density (i.e, an order of magnitude smaller than the critical initial density we calculate in this section) to the other WD in the direction of the relative motion. With this definition of the interaction area, a layer with the critical density for detonation must face another layer with the same order of magnitude of the density. If the region directly interacting in one WD is too diluted, it will not compress the high-density region of the interacting WD, thereby preventing it from reaching the critical density needed for Carbon detonation. Therefore, we make sure that a shock (from the collision) will be produced and compress this layer as a result. In case there is no direct collision between layers with similar order of magnitude densities, ignition is impossible, as there will be no compression and heating in dense enough material to allow for a nuclear runaway. Instead, the high-density region would simply move through the low-density one. As can be seen in Table~\ref{tab:simulations_parameters}, this implies that since the direction of motion changes slightly before the moment of impact, so does $\rhoint$. Here, we aim to identify the minimal initial value of $\rhoint$ that can lead to a post-shock (from the collision) density above the critical value that leads to Carbon detonation.  This is done while considering the result of the direct collision in the interaction region (between the two purple, partial dashed lines in Fig. \ref{fig:sketch_initial_configuration}).
Here, we only consider the properties of the WD resulting from the pure collision to determine if a detonation is a possible outcome. However, when these indeed lead to a detonation, the following shock will change the final properties of the system as calculated by \cite{2014KushnirKatzSelfSimilarHeadOnSolution}.

Since the upstream pressure $p_1$, $\mu\equiv\frac{m^2}{\rho_1 p_1}$ is always positive for a given mass $m$ going through the shock,  the asymptotic Hugoniot curve goes to:
\begin{equation}
    \Tilde{v} \equiv \frac{\rho_1}{\rho_2} \geq  \frac{\gamma - 1}{\gamma + 1} ,\\
    \rho_2 \geq \rho_c \\
    \Rightarrow \rho_1 \geq \frac{\gamma - 1}{\gamma + 1} \rho_c 
\end{equation}
Where 1 and 2 subscripts stand for upstream and downstream of the shock and $\gamma$ is the specific heat ratio. 
For an ignition density of $\rho_c=3\cdot {10}^6 \unit{\gcms} $\citep{ryan2010stellarBook} and $\gamma=5/3$ (where in reality it can be lower to a minimum of $\gamma=4/3$, leading to larger compression and resulting in a lower critical $\rho_1$):
\begin{equation}
\label{eq:density}
    \rho_1 \geq \frac{1}{4} \rho_c = 7.5 \cdot {10}^5 \unit{\gcms}
\end{equation}
Therefore, for this case, the minimum value of the density in the interaction layer, $\rhoint$, at the moment of impact is $\rhoint =  \rho_1 \geq  7.5 \cdot {10}^5 \unit{\gcms}$. 
We now find the criterion for the minimal (critical) relative velocity upon collision such that a detonation is formed. The shock velocity for the detonation should be the Chapman-Jouguet velocity, which is derived at the upper Chapman-Jouguet point, where the Rayleigh line is tangent to the Hugoniot curve, which is also the sonic point of the downstream, i.e., $u_2 = c_s$.  From the conservation of mass:
\begin{equation}
    \rho_1 u_1=\rho_2 u_2 = \rho_2 c_s = \rho_2\sqrt{\gamma \text{R} T_2} \geq \rho_2\sqrt{\gamma \text{R} T_c}
\end{equation}
where R is the gas constant. For a detonation $\frac{\rho_1}{\rho_2}<1$, therefore:
\begin{equation}
    u_1=\frac{\rho_2}{\rho_1}\sqrt{\gamma \text{R} T_2} \geq \sqrt{\gamma \text{R} T_2} \geq \sqrt{\gamma \text{R} T_c}
\end{equation}
and since the critical temperature for Carbon detonation is $T_c\approx1.3\cdot{10}^9 \unit{\text{K}}$ (together with $\gamma=5/3$ consistent with the EOS for the WD shown below) we get:
\begin{equation}
    u_\text{relative}\geq4.24\cdot {10}^8 \unit{\text{cm}/\text{s}},
\end{equation}
which is the case for all of our simulations at the moment of impact.
We note that the exact temperature for Carbon ignition is not resolved in the literature, and it depends on the density. The relevant temperature range is between $5\cdot {10}^8\unit{\text{K}}$, at which nuclear burning begins, and detonation ignition considered up to $3\cdot {10}^9 \unit{\text{K}}$. Here, we used the value of $1.3\cdot{10}^9 \unit{\text{K}}$, consistent with this range, and providing an excellent fit to the results of our simulations, as well as past relevant simulations we present here in Tab. \ref{tab:past_simulations}. This value depends on $\gamma$, which, as mentioned, can be lower than the value we used here, however, not significantly. For the model used here, of $0.8 \text{M}_\odot$ CO WD, $\gamma=5/3$ works well, but a lower gamma is more appropriate for higher mass WDs.

Furthermore, with our criteria, as in reality, one can have an explosion with a lower density as long as the temperature is sufficiently high. The actual critical values of detonations are not independent of each other and are not certain in the literature, however, one can always use our criteria with different (perhaps updated) values. Here, we want to determine if the system will explode or not, and we require both criteria (density and temperature) to be fulfilled together. This will lead to not only an ignition but a detonation wave, full burning, and an explosion. As we show in the button panels of Fig. \ref{fig:combo}, when the temperature reaches the critical value but the density does not, there is no detonation and thus no explosion. This will also be the case if only the density reaches the critical value. In Table. \ref{tab:past_simulations}, we show how these criteria apply to previous simulations. One can notice this dependence on the relative velocity and $\rhoint$ when comparing the different simulations with non-zero impact parameters by \cite{2009RaskinWDCollisionSimulation,2010RaskinNi56FromWDCollisions}, resulting in a remnant in the low mass cases and in a detonation for the higher masses with the larger impact velocity and higher maximum density at the collision area.

In addition to these criteria, one can verify that the conditions hold in a minimal volume containing the detonation point according to \cite{2006DursiTimmesIgnitionSize}.
We note that in all of our simulations, the temperature does not change significantly before the actual collision (as claimed by \citealt{2014KushnirKatzSelfSimilarHeadOnSolution}), but it only changes after the detonation occurs and due to the shock compression.

We can use these criteria to calculate the impact parameter $b$, for which the highest density in the interaction layer, $\rhoint$ is higher than the critical density calculated above and test this in our detailed simulations.
In future work, one can calculate the initial impact parameter, $b^i$ (marked as $b$ in Fig. \ref{fig:sketch_initial_configuration}), such that the impact parameter at the moment of impact due to gravitational focusing, leads to $\rhoint$ at the moment of impact to be larger than the critical value calculated above.

\section{Methods}
\label{methods}
\begin{figure}
    \centering
    \includegraphics[width=\linewidth,clip]{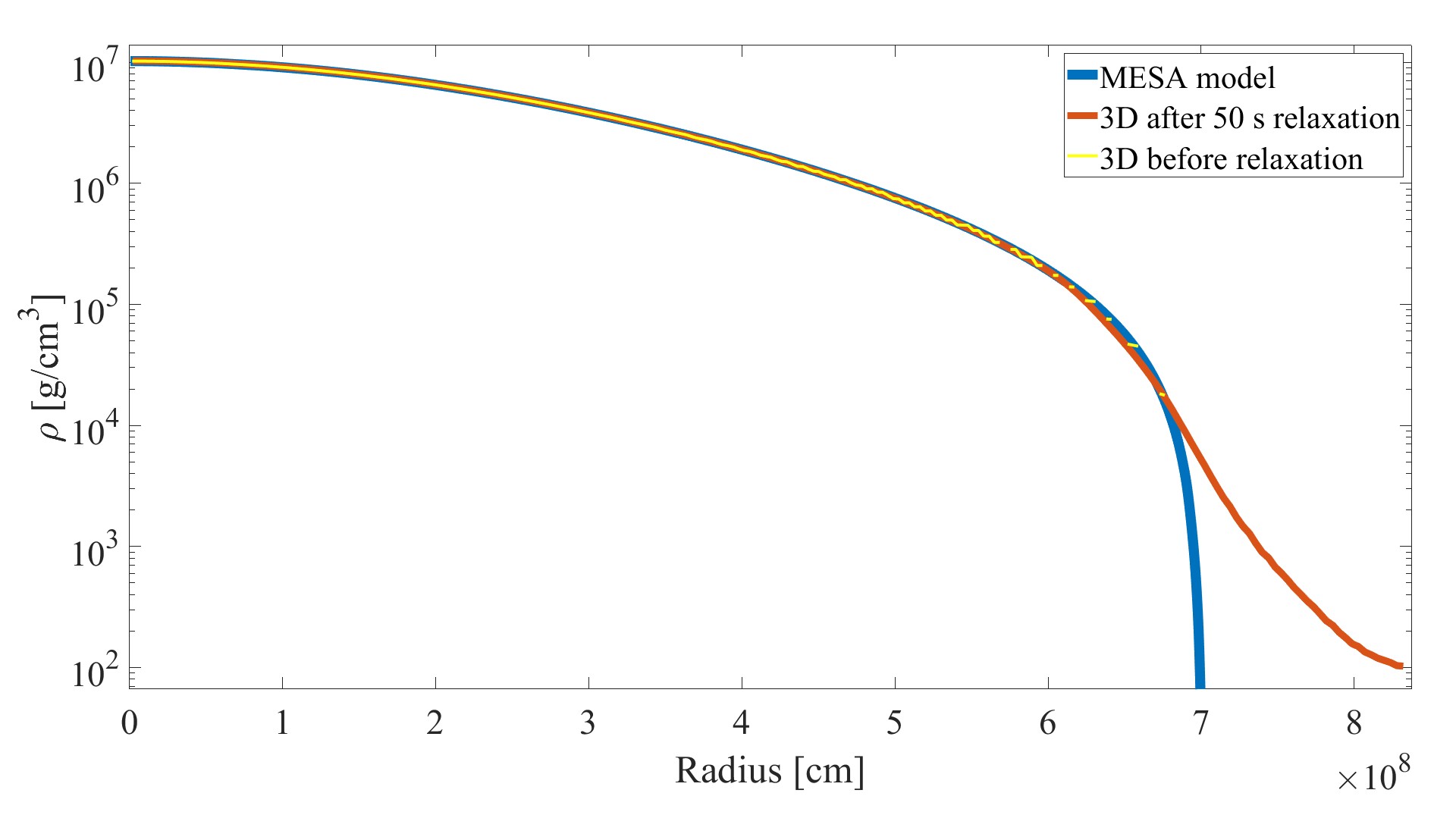}
    \caption{The density profile of the initial WD model produced with the 1D stellar evolution code \texttt{MESA}, compared with the generated 3D profiles before and after the relaxation stage. }
    \label{fig:initial_density}
\end{figure}

We simulated the collision of two equal-mass CO WDs with 5 different impact parameters using the moving-mesh magneto-hydrodynamical code \texttt{AREPO} \citep{2010SpringelArepo, 2020WeinbergArepoPublic,2011PakmorArepoMHD}. 
    To create our initial binary components, we first use the 1D stellar evolution code \texttt{MESA} \citep{2011MESA, 2013mesa, 2015mesa, 2018mesa, 2019mesa, 2023mesa} to create the initial model of a $0.8 \text{M}_\odot$ white dwarf (see density profile in Fig. \ref{fig:initial_density} and more details in \citealt{2021PakmorZenatiHybridIgnition} ). 
We then create 3-dimensional initial conditions for \texttt{AREPO} \citep{2010SpringelArepo,2020WeinbergArepoPublic}. We generate the 3D mesh with healpix shells \citep{2012PakmorSPHTypeIa, 2017OhlmannStableArepoModel}, with the same method as described in \cite{2021PakmorZenatiHybridIgnition}. In addition to the movement of the mesh generating points, cells in our simulations are refined and eliminated when their mass is more than a factor of two away from the target gas mass, which we chose to be $m_\text{target}=2\cdot{10}^{26}$g, and are always refined if their volume is more than 10 times larger than their largest direct neighbor. We include the self-gravity of the gas, with a gravitational softening of the cells $~2.8$ times their radii, with a minimum softening of $10$ km. The degenerate electron gas of the WDs is modeled using the \texttt{Helmholtz} equation of state \citep{2000TimmesHelmholtzEOS} with Coulomb corrections. The minimum volume of a cell in the simulation is about $\left(14 \text{ km}\right)^3$ (at the maximum compression), and the maximum is $~\left(25 \cdot 10^4 \text{ km}\right)^3$, from a total size of $\left(10 ^7 \text{ km}\right)^3$.

\begin{figure}
    \centering
    \includegraphics[width=\linewidth,clip]{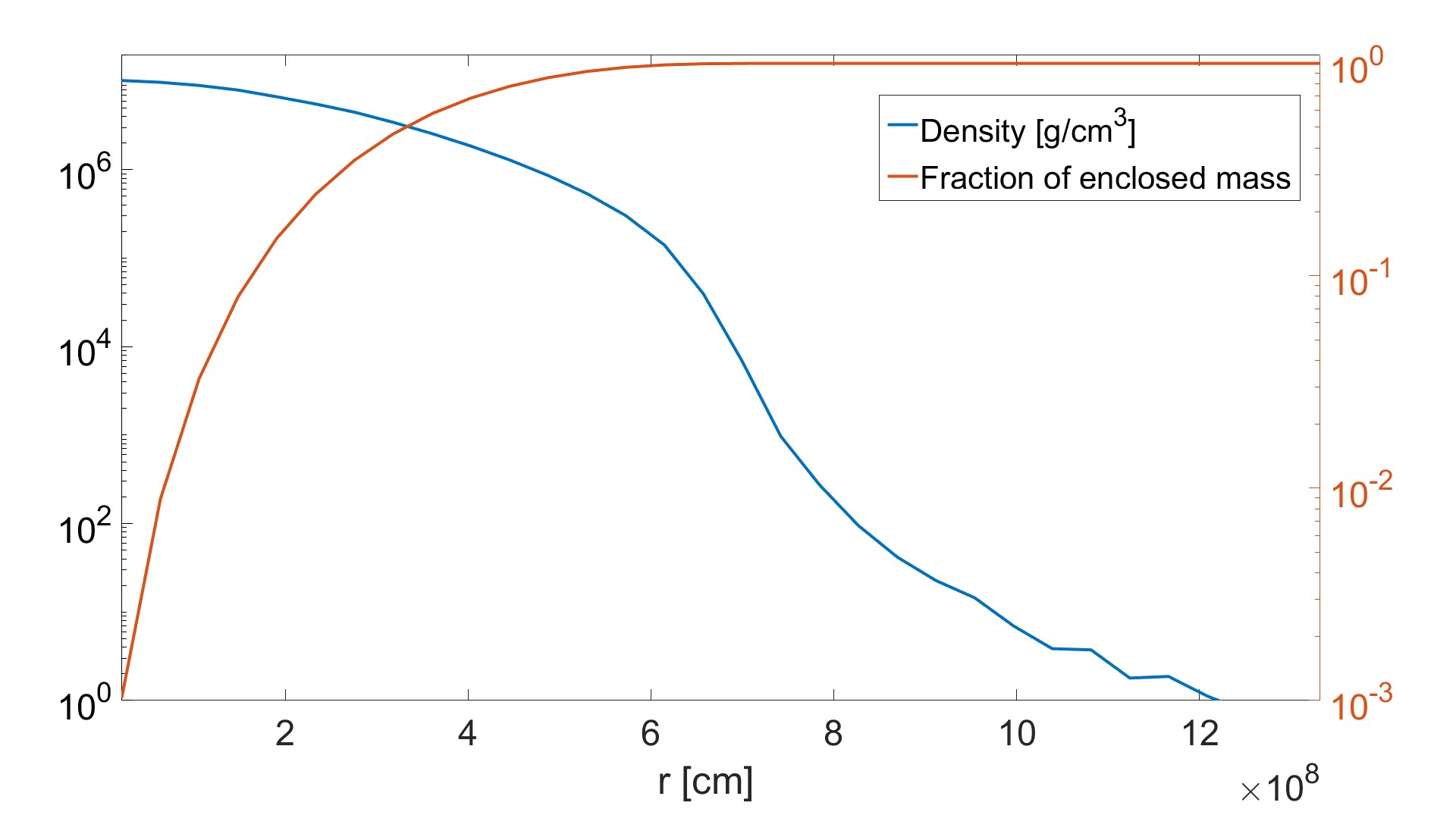}
    \caption{The radial density profile of the initial WD model after the relaxation stage, and the corresponding fraction of enclosed mass.}
    \label{fig:initial_cum_mass-density}
\end{figure}

\begin{figure}
    \centering
    \includegraphics[width=0.7\linewidth]{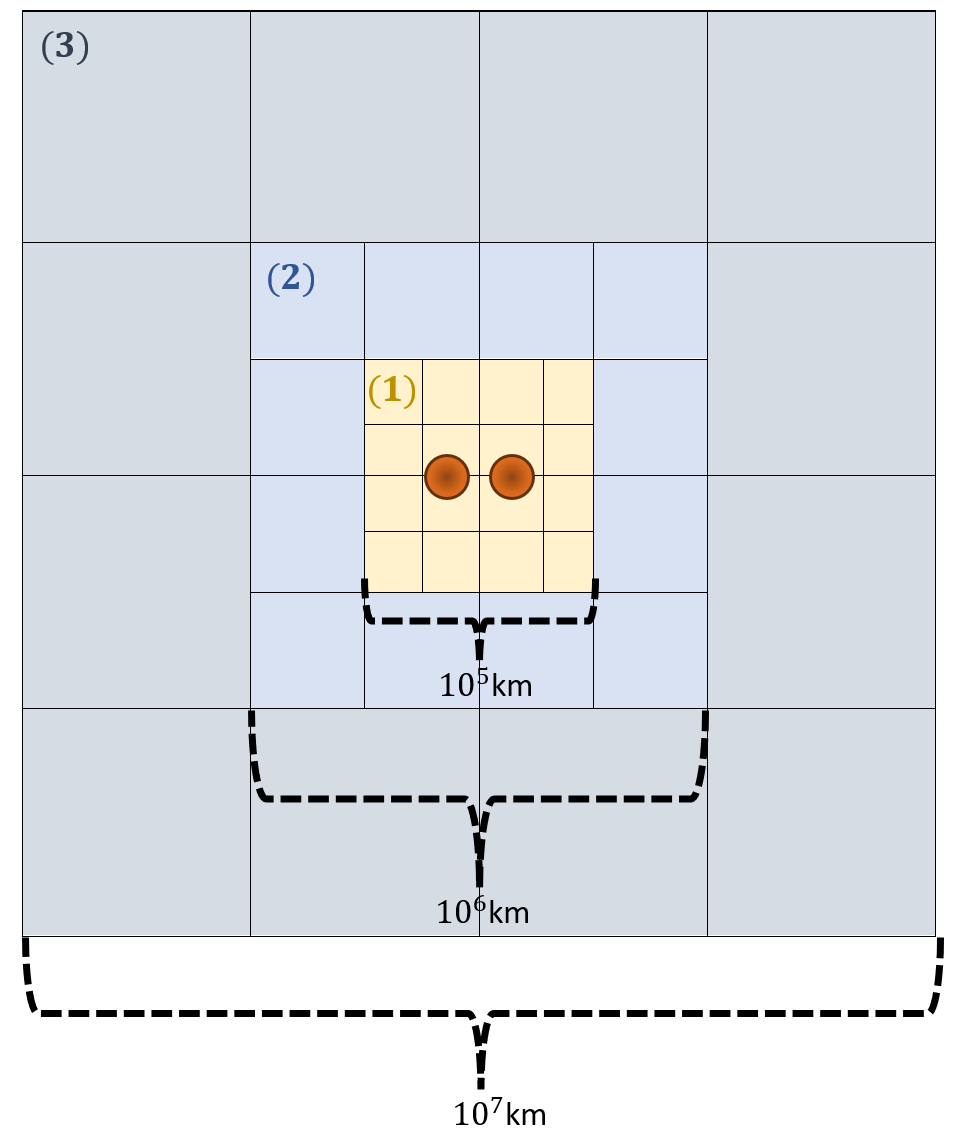}
    \caption{Demonstration of the cells in our grid, with the innermost box around the two WDs (area (1)) with 32x32x32 background cells, another grind around it (area (2)) with 32x32x32 cells up to ${10}^6$ km, and finally the largest box with 32x32x32 cells with the size of ${10}^7$ km. We note that for simplicity we show only 4 cells per dimension in this sketch instead of the 32. In addition, after the start of the simulation, the mesh is refined such that the difference between the volume of two neighbor cells is not more than 10 (where the larger cells are refined.}
    \label{fig:nested_grid}
\end{figure}

In order to avoid numerical noise from the code and coordinate transitions, we perform a relaxation stage for about $T_\text{max}=50$s, in which we let the velocity field evolve adiabatically. We use the same method as described in \cite{2012PakmorSPHTypeIa}, where we divide this dynamical stage into three parts: (1) during the first 20 percent we damp the velocities and momenta by a factor of 10 times the size of the current time step, (2) we then gradually decrease this factor until it reaches 0.01, and then (3) stop the damping for the last 20 percent and let the model evolve freely in isolation. These velocity reductions can be written as follows:
\begin{equation}
    \text{d} v \left(t\right) = 
    \text{d}t \cdot 10 \cdot \begin{cases}
    1 & \frac{t}{T_\text{max}} < 0.2 \\
    {10}^{5 \left(0.2-\frac{t}{T_\text{max}}\right)} & 0.2 < \frac{t}{T_\text{max}} < 0.8\\
    0 & \frac{t}{T_\text{max}} > 0.8
    \end{cases}
\end{equation}
We verify the stability of this model by comparing the density profile of our model during the different times of the relaxation stage and allowing for a maximum change of less than one percent in Fig. \ref{fig:initial_density}. 
After relaxation, we place the binary components at a distance of about the Roche-lobe radius \citep{1983eggleton} in the plane of orbital motion, where we calculate the radius of the WD to be the point where the density drops below a single $\gcms$, which is at $R_{WD}\approx 1.26\cdot {10}^9\cm$ (see Fig. \ref{fig:initial_cum_mass-density}).
We note that since most of the mass is below half this calculated radius (see Fig. \ref{fig:initial_cum_mass-density} our initial separation is effectively more than a factor of two larger than the actual Roche-lobe size and is further increased for cases of nonzero impact parameters. These models are initially surrounded by nested grids with a decreasing resolution- each with 32-cubed (initial) cells, from ${10}^{5}$ km that encloses the initial system, to ${10}^{7}$ km to accommodate the ejecta of from the explosion for the time of the simulation. We set the background density to ${10}^{-5}\gcms$. We demonstrate this initialization in Fig. \ref{fig:nested_grid}.

We use a 55 isotope nuclear reaction network, coupled to the hydrodynamics as described in \cite{2012PakmorSPHTypeIa}. Nuclear reactions are calculated in all cells, except those inside the shock front, following \cite{2009SeitenzahlSpontaneousDetonationWD}, when $\nabla\cdot \Vec{v}<0$ and $\left|\nabla P\right|r_\text{cell}/P > 0.66$  with the JINA reaction rates \citep{2010CyburtJinaReaclib}. 

The initial vertical distance, or the impact parameter, varies from 0 (the case of a head-on collision), to 2.5 times $R_{WD}$. We present a schematic graph of this initial configuration in Fig. \ref{fig:sketch_initial_configuration}. We note that our definition of the radius of the WD, $R_{WD}$, is as described above, and one can use a different density threshold (usually larger) to define the radius. 

We initialized each of the WDs with a magnetic dipole along the z-axis, following the same approach as in \citep{2016OhlmannMHD} and followed the magnetic evolution during the simulation. However, the low magnetic fields used are irrelevant dynamically by themselves, and we find they play no role in our simulations.

%

\begin{figure}
    \centering
    \includegraphics[width=\linewidth,clip]{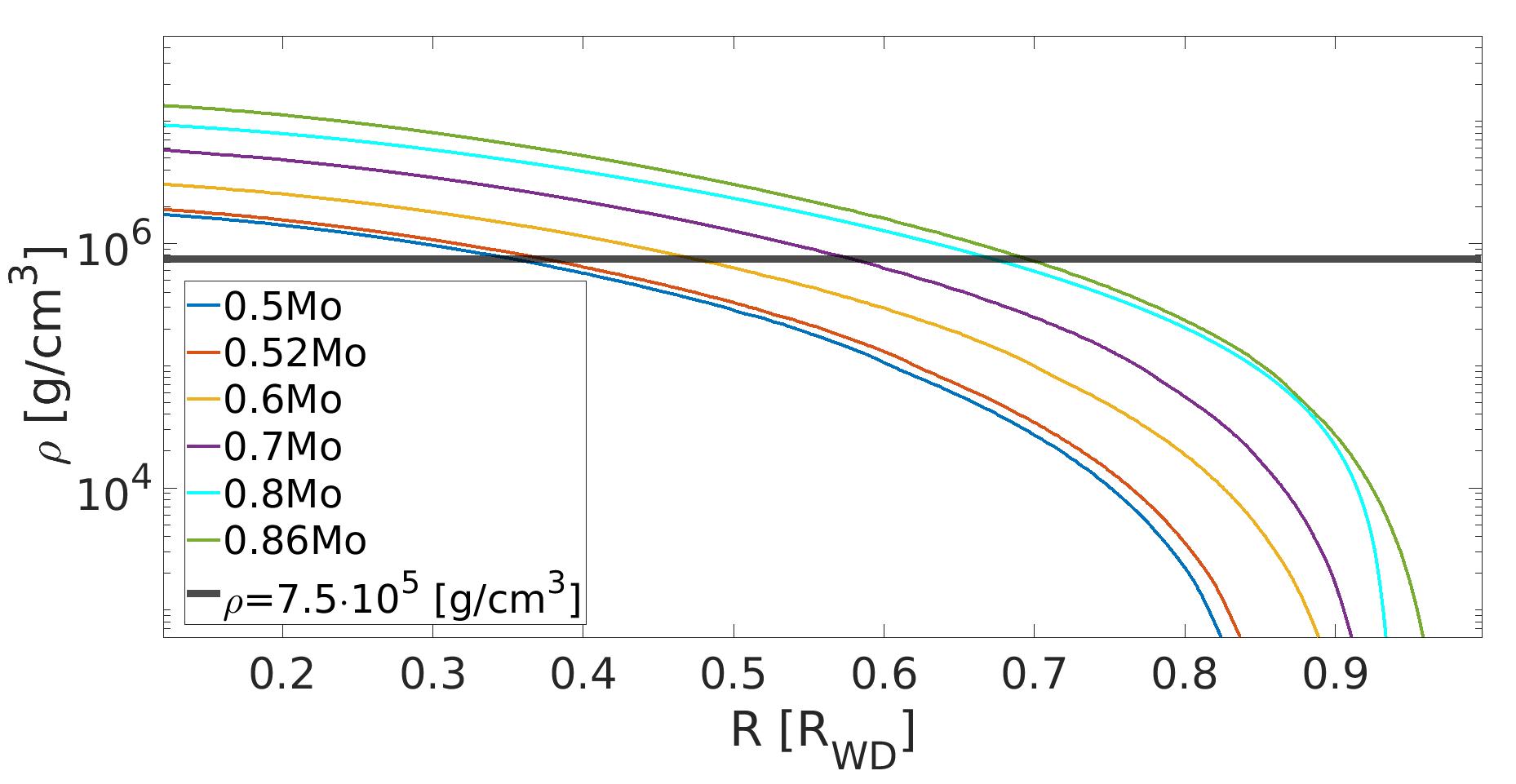}
    \includegraphics[width=\linewidth,clip]{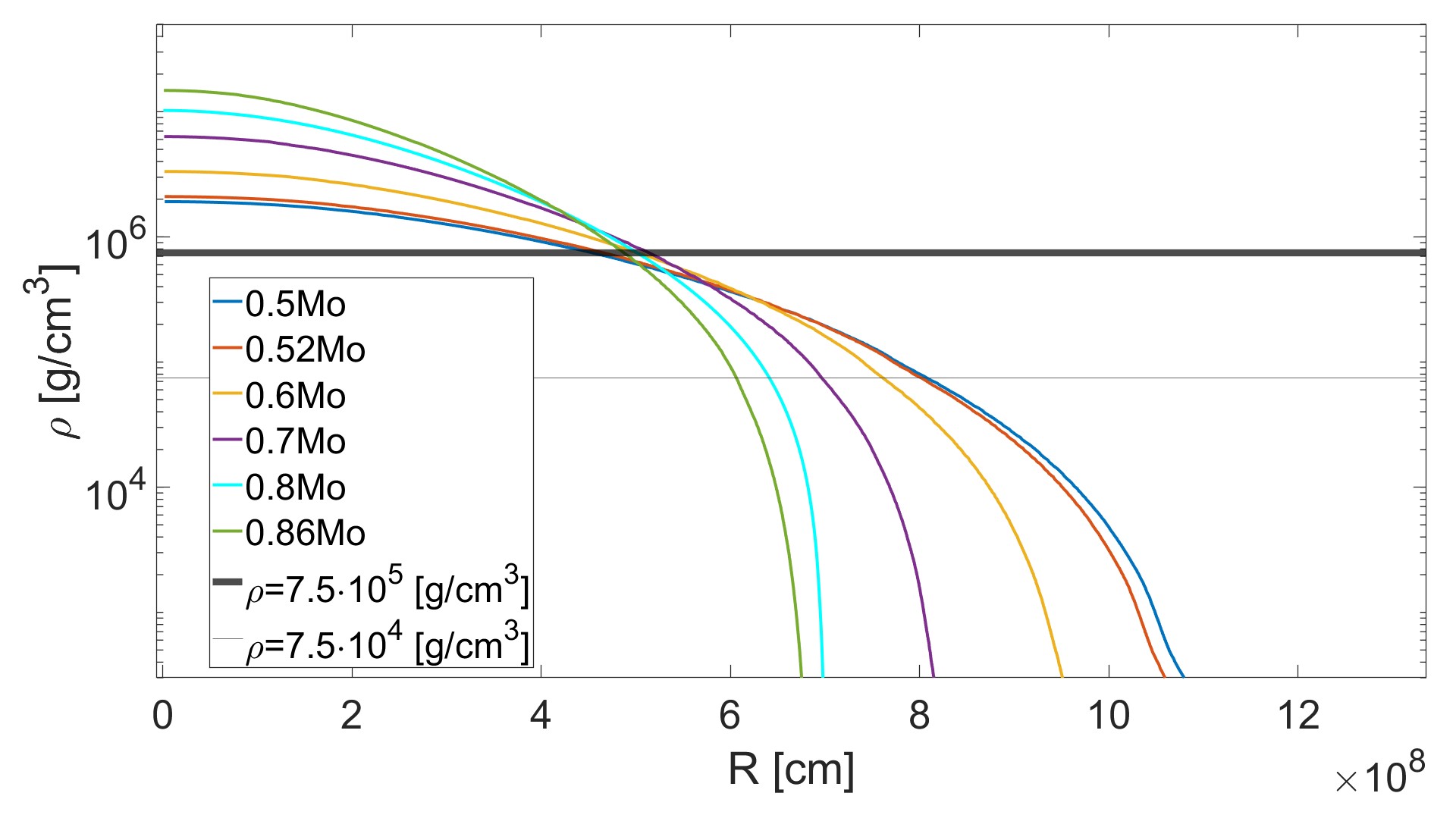}
    \caption{The radial density profile of different WD models, showing the critical distance $d$ of the interacting layer (which implies an impact parameter $b=R_{WD} + d$) to allow for the explosion.  The upper panel shows the distance in units of the WD radius. Bottom: distance in cm.}
    \label{fig:density-r}
\end{figure}

Fig. \ref{fig:density-r} shows that the critical density $\rhoint$ calculated in Sec. \ref{explosion criteria} is achieved in approximately the same radius, $d$, for different CO WD masses, where it encloses a larger fraction of the mass for more massive WDs. Therefore, for equal mass CO WD collisions, one can simply find the values of the impact parameter, $b$, and the relative velocity, $u_\text{relative}$, at the moment of impact from gravitational focusing and compare them with the critical values found above, while the critical value for the impact parameter at this moment is $b=d+R_{WD}^{10c}$, where $R_{WD}^{10c}$ is the distance of the layer with an order of magnitude smaller density than the critical density we calculated in Eq. \ref{eq:density}.

\section{Results}
Overall, we tested the collisions of the two equal WD models with 10 different initial vertical separations and three initial velocities. The different impact parameters correspond to different densities in the interacting region, $\rhoint$ (using $b=d+R_{WD}^{10c}$ and Fig. \ref{fig:density-r}), therefore were chosen to cover the different regimes needed to verify our calculated criteria in Sec. \ref{explosion criteria}. We expect typical collisions to be dominated by the escape velocity from the WD, but we also considered much higher velocities to explore the dependence of the results on the collision velocity, as well as to test our analytic expectations vis-a-vis the detailed models under varied conditions. In addition, high velocities are possible in nature in extreme regions, such as close to supermassive black holes in galactic nuclei, where hypervelocity collisions can occur. In total, we tested our calculations by performing 23 different simulations.{} We present the outcome and production of ${}^{56}$Ni in Tab. \ref{tab:simulations_parameters}, and the \Ni{} results of our numerical configuration tests in Tab. \ref{tab:simulations_parameters_resolution}. 

We began by simulating the case of a rapid collision with relative velocities of $1.6\cdot {10}^9 \text{cm}/\text{s}$, where we find a strong correlation between the highest density in the interacting region ($\rhoint$ in Fig. \ref{fig:sketch_initial_configuration}) and the production of ${}^{56}$Ni, the detailed results are presented in the following subsection. We later tested two cases of collisions with a smaller initial velocity, equal to the terminal velocity presented in \ref{sec:free_falling}, and one case of a collision with the threshold impact parameter and velocity in between the two.
We also verified that the temperature and the density at the detonation point just before ignition were very close to the values we assumed in our calculation at Sec. \ref{explosion criteria}.
\begin{table*}
    \centering
    \begin{tabular}{|c|c|c|c|c|c|c|}
    \hline
    \hline
     Simulation & $\rhoint$ & impact parameter & impact velocity & outcome & ${}^{28}$Si & ${}^{56}$Ni   \\
    Name & $\left[\text{g}/\text{cm}^3\right]$ & $\left[\text{cm}\right]$ & ${10}^{8}\left[\text{cm}/\text{s}\right]$ & & $\left[\text{M}_{\odot}\right]$ &$\left[\text{M}_{\odot}\right]$ \\
    \hline
    \hline
     H0 & $1.1\cdot {10}^7$ & 0 & 16 & explosion & 0.233 & $0.937$\\
     HS & $6.7\cdot {10}^6$ & $8.39\cdot {10}^8$ & 16 & explosion & 0.494 & $0.196$\\ 
     HC & $9.8\cdot {10}^5 (8\cdot {10}^5)$ & $1.05\cdot {10}^9 (1.08\cdot {10}^9)$ & 16 & explosion & 0.466 & $0.185$ \\ 
     HLC & $6.2\cdot {10}^5 (5\cdot {10}^5)$ & $1.1\cdot {10}^9 (1.12\cdot {10}^9)$ & 16 & no explosion & ${10}^{-4}$ & 0 \\ 
     HL & $6.9\cdot {10}^4 (-)$ & $1.21\cdot {10}^9 (1.26\cdot {10}^9)$ & 16 & unbounded WDs, no explosion & 0 &  0 \\
    \hline
     2FFC & $2.75\cdot {10}^6 (8\cdot {10}^5)$ & $9.62\cdot {10}^8 (1.08\cdot {10}^9)$ & 8.2 (7.14) & explosion & 0.429 & $0.344$ \\
    \hline
     FF0 & $1.1\cdot {10}^7$ & 0 & 5.8 (3.57)& explosion & $0.487$ & $0.267$ \\
     FFS & $9.2\cdot {10}^6 (2.1\cdot {10}^6)$ & $7.84\cdot {10}^8 (9.9\cdot {10}^8)$ & 5.48 (3.57) & explosion & $0.476$ & $0.282$ \\ 
     FFM & $6.3\cdot {10}^6 (8\cdot {10}^5)$ & $8.5\cdot {10}^8 (1.08\cdot {10}^9)$ & 5.3 (3.57) & explosion & $0.486$ & $0.251$ \\ 
     FFML & $ 2.9 \cdot {10}^6 (7.52 \cdot {10}^4)$ & $ 9.82 \cdot {10}^8  (1.28\cdot {10}^9)$ & 5.17  (3.57) & explosion & $0.475$ & $0.272$ \\ 
     FFL & $7.5 \cdot {10}^4 (-)$ & $1.2\cdot {10}^9 (1.72\cdot {10}^9)$ & $4.4$ (3.57) & unbounded WDs, no explosion & 0 & 0\\
    \hline
    \hline
    \end{tabular}
    \caption{Initial parameters and results for the different simulations, where there are two $0.8 \text{M}_\odot$ CO WDs with radii $R_{WD} \approx 1.26\cdot {10}^9 \text{cm}$ (determined to be the position at which the density drops below one $\gcms$). The first columns show the initial configurations, while the last 3 are the final outcome and final mass production of ${}^{28}$Si and ${}^{56}$Ni. $\rhoint$ is the maximum density of the WDs along the region of interaction in the direction of motion (the x-axis in our case), which is illustrated in Fig. \ref{fig:sketch_initial_configuration}. Values in parentheses are those at the initial distance. We define the outcome either as a full explosion, i.e., a complete unbinding of the WDs, or no explosion (though a limited burning may be seen) where a bounded remnant is left. In case of remaining unbounded WDs, the two survivours will not be bound in a binary system.}
    \label{tab:simulations_parameters}
\end{table*}

\subsection{Hypervelocity Collisions}
We simulated the collisions of the two WDs with an initial relative velocity of $1.6\cdot {10}^9 \text{cm}/\text{s}$, which may likely occur only close to supermassive black holes, where the relative velocities are dominated by the orbital velocities around the supermassive black hole. We considered five different initial vertical separations, starting with head-on collisions with a zero impact parameter up to collisions at an initial vertical separation of $1.26\cdot {10}^9 \cm$ (see Tab. \ref{tab:simulations_parameters}). In the following, we discuss the different cases.

\begin{figure*}
    \centering
    \includegraphics[width=\linewidth,clip]{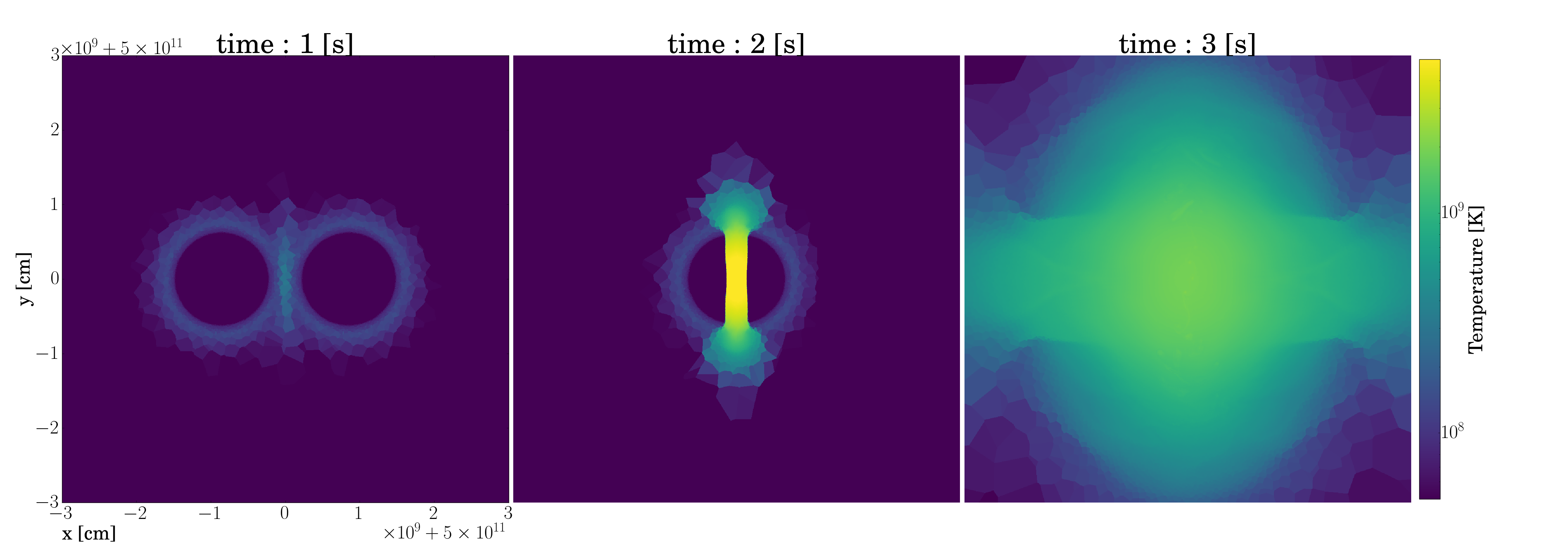}
    \includegraphics[width=\linewidth,clip]{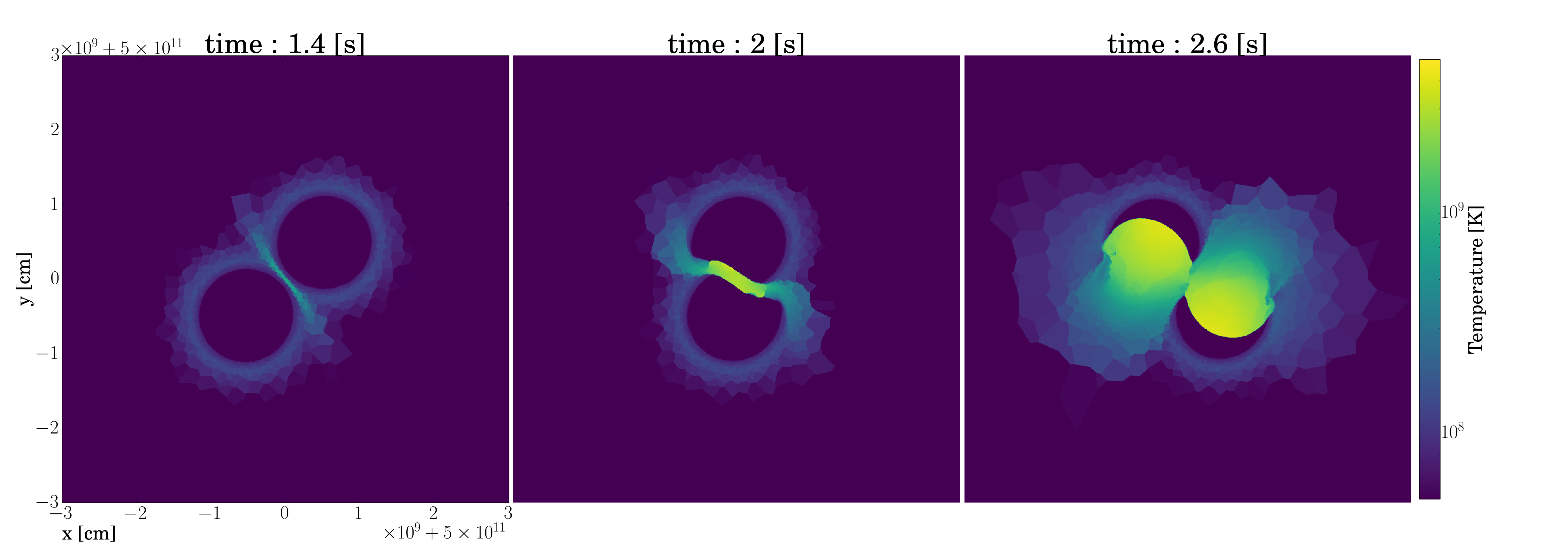}
    \includegraphics[width=\linewidth,clip]{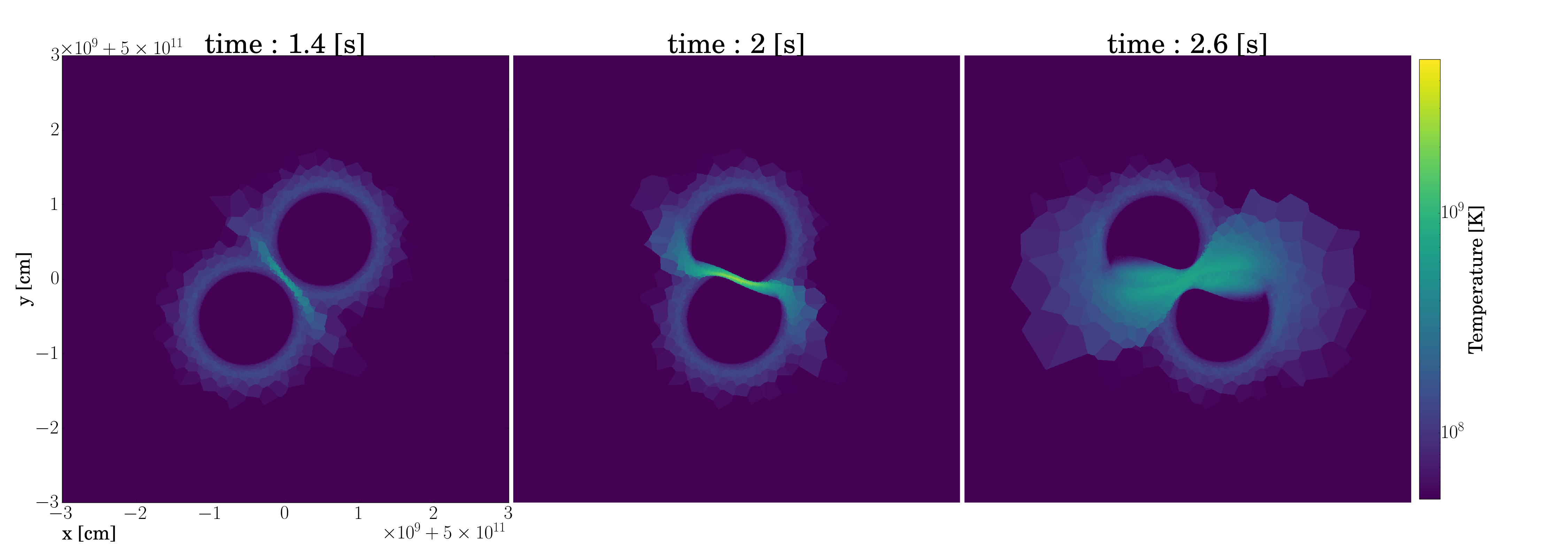}
    \caption{Top panel (case H0- Tab. \ref{tab:simulations_parameters}): Head-on collision with an initial relative velocity of $1.6\cdot {10}^9 \text{cm}/\text{s}$ at different times- before colliding, during the detonation and shortly after the explosion.
    Mid panel (case HC): The case of a collision with an initial velocity of $1.6\cdot {10}^9 \text{cm}/\text{s}$, and an impact parameter corresponding to interacting density comparable with the critical density calculated in Sec. \ref{explosion criteria}. Bottom (case HLC): The collision with an impact parameter slightly larger than the critical one calculated following the critical interacting density from Sec. \ref{explosion criteria}, producing only a small amount of ${}^{28}$Si, but no \Ni{} and no explosion. }
    \label{fig:combo}
\end{figure*}

\subsubsection{Grazing collision- case HL} 
In the case of a grazing collision, we expect the maximum density in the interaction layer to be $\rhoint = {10}^5 \gcms$, much too low for ignition.
In this case, the WDs were only slightly deformed but showed no significant interactions, and in particular, we observed no Carbon ignition.


\subsubsection{Head-on collision- case H0}
The simulation with the energetic head-on collision was expected to lead to ignition, and it indeed leads to a detonation and produced the highest amount of \Ni, with more than half of the mass in the system converted into \Ni. In the top panel of Fig. \ref{fig:combo}, we plot the temperature slices of this scenario before the WDs touch each other, during the burning and the beginning of the explosion, and shortly after the explosion. In Subsec. \ref{sec:res}, we present another run with different numerical configurations (as in \citealt{2013KushnirKatzHeadOnWDCollisionsInTriple}) that produced similar results to the case discussed here.

\subsubsection{Collision with a small impact parameter- case HS}
In order to get another clear non-head-on explosion case, we chose a small impact parameter such that the two WDs overlap by two-thirds of their volume, with $b=2/3 \cdot R_{WD}$. This case leads to a detonation, leading to a smaller ${\rm ^{56}Ni}$ mass being produced.

\subsubsection{Collision with critical density at the interacting surface- case HC}
To test the critical impact parameter criteria we found in Sec. \ref{explosion criteria}, we simulate a collision with an initial vertical separation of $1.12 \cdot {10}^9 \cm$, which corresponds to a maximum density at the interacting layer to be slightly larger than the critical threshold density, with $\rhoint=8\cdot {10}^5 \gcms$ (see Fig. \ref{fig:density-r}). In this case, we indeed got an explosion with less \Ni{} production than in the head-on case. We present some snapshots of this scenario in the middle panel of Fig. \ref{fig:combo}.

\subsubsection{Collision with slightly larger impact parameter than the threshold- case HLC}
Continuing the testing of the criteria from Sect. \ref{explosion criteria}, we simulate the collision with an initial vertical separation of $1.1 \cdot {10}^9 \cm$, which corresponds to a maximum density at the interacting layer of the threshold density, with $\rhoint^i=5\cdot {10}^5 \gcms$, and $\rhoint=6.2\cdot {10}^5 \gcms$ at the moment of impact, which is slightly below the calculated threshold (see Fig. \ref{fig:density-r}). In this case, carbon is ignited and burned to ${}^{28}$ Si, but only gives rise to incomplete burning, without a detonation (the density cannot be compressed to above the critical density) and thus no further burning is found, nor production of \Ni{} and no explosion leading to the unbinding of the WDs. Some snapshots of this case are presented at the bottom of Fig. \ref{fig:combo}.
A shock occurs in the interaction region due to the direct impact of the supersonic relative motion of the WDs. Consequently, this region shows compression and heating, leading to nuclear burning of Carbon. Nevertheless,  unlike explosive cases, the shock does not propagate supersonically and remains localized only in the interaction surface. We note that no explicit heat conduction is used in our simulations.

\subsection{Collisions between free-falling objects}
\label{sec:free_falling}
We consider 4 cases of collisions with a terminal velocity $u_\text{relative} = \sqrt{2G\left(m_1+m_2\right)/\Delta x}$, where in our case $\Delta x = a_\text{RL}\approx 3.32\cdot 10^9 \cm$, as described in Sec.\ref{methods} and demonstrated in Fig. \ref{fig:sketch_initial_configuration}. This corresponds to a collision at the terminal (escape) velocity of the WDs from a free-falling motion.

\begin{figure*}
    \centering
    \includegraphics[width=0.45\linewidth,clip]{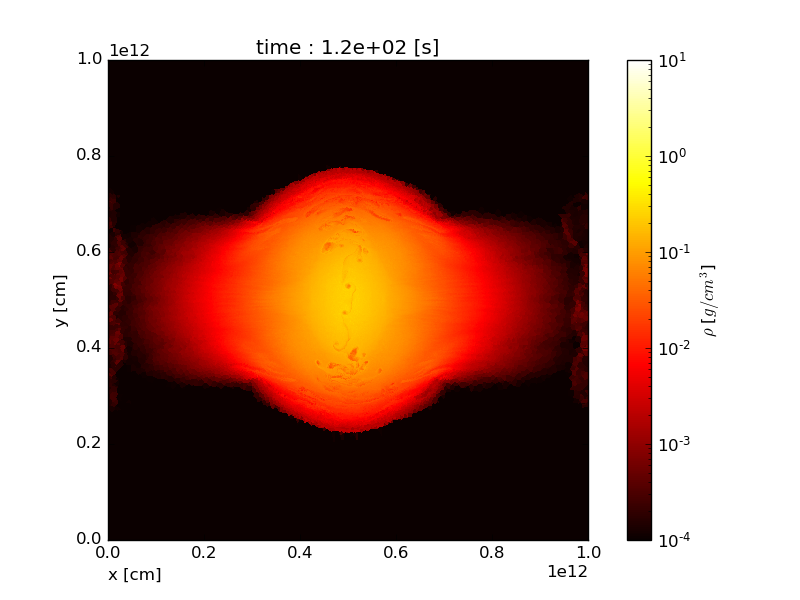} \includegraphics[width=0.45\linewidth,clip]{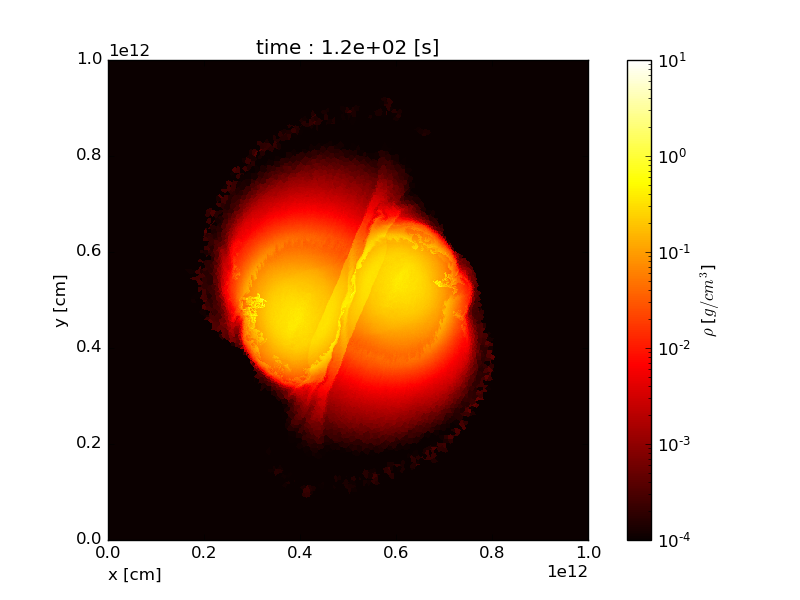}
    \includegraphics[width=0.45\linewidth,clip]{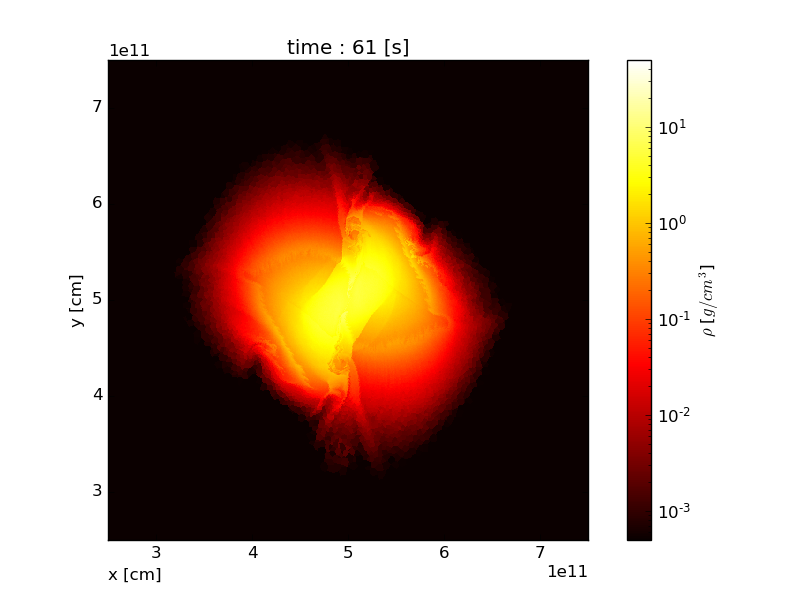} \includegraphics[width=0.45\linewidth,clip]{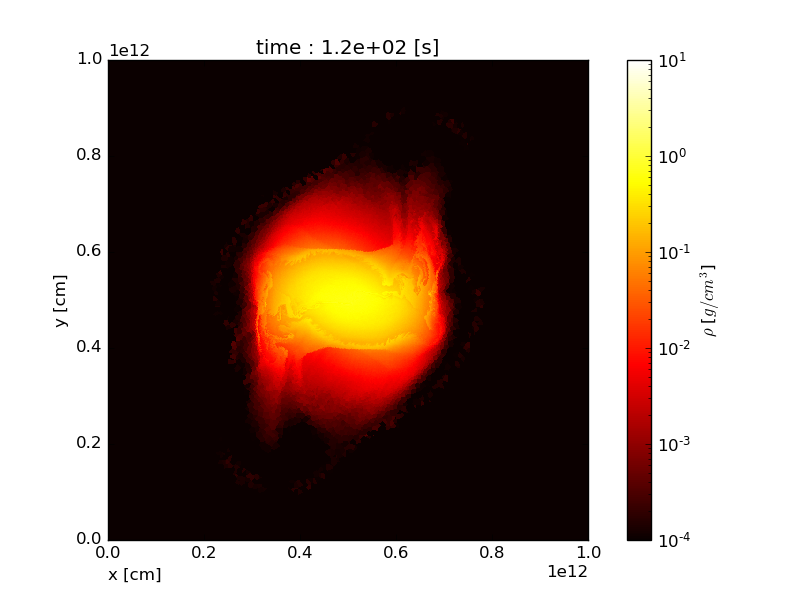}
    \caption{Density slices of the last snapshot for different simulations, showing the ejecta from the explosion. From top left, clockwise order: Head-on collision at hyper-velocity (H0 in Tab. \ref{tab:simulations_parameters}); threshold collision at hyper-velocity (HC); threshold collision at terminal velocity (FFS);  and threshold collision at twice the terminal velocity (2FFC).} 
    \label{fig:ejecta}
\end{figure*}

In the limiting case where $\rhoint=8\cdot {10}^5$ (case FFS), the actual vertical distance upon collision is smaller than the initial one (as we show in Tab. 
\ref{tab:simulations_parameters}, where original values, if different than values upon impact, are in parenthesizes), since in that case, the velocity is sufficiently small, as to allow for gravitational focusing, therefore, we expect the effective $\rhoint$ to be larger than $8\cdot {10}^5$ (see Fig. \ref{fig:density-r}). The ejecta of this case is presented in Fig. \ref{fig:ejecta}. 
We ran another simulation (case FFM) with the same initial impact parameter as case HLC, which only produced ${}^{28}$Si in the hypervelocity case but with the initial free-fall velocity. In this case, FFM (Tab. \ref{tab:simulations_parameters}), the significant reduction of the impact parameter resulted in an explosion. To verify our results near the calculated boundary, we ran another simulation (FFML).  Due to gravitational focusing, this simulation resulted in a smaller impact parameter, but one still close to the limiting case (HLC), leading to a full explosion as expected.

In addition, we ran the head-on case (FF0) for a direct comparison with the hypervelocity head-on case (H0). This was also later explored  (see Subsec. \ref{sec:res}) with different numerical configurations. When we compare the results of \Ni{} production for the same numerical configuration but for different initial velocities, we find that the \Ni{} abundance produced by the explosions decreases with lower initial velocities. We further discuss the effect of the different numerical configurations in Subsec. \ref{sec:res}.
The last scenario we present here for the free-falling case is the large initial vertical separation- case FFL, which had $\rhoint=1.2\cdot 10^5 \gcms$ upon impact, i.e- below the critical calculated $\rhoint$ (see Fig. \ref{fig:density-r}), and did not explode.

\subsection{Collisions with a medium initial velocity (case 2FFC)}
We run only one simulation with an intermediate velocity of twice the free-fall velocity of the WDs and with an impact parameter just below the detonation threshold. We find that a detonation occurs, as expected, producing 0.34 M$_\odot$ of \Ni{}, which is higher than both the high-velocity runs (cases HC, HS in Tab. \ref{tab:simulations_parameters}) and the escape velocity runs (cases FFS, FFM), which gave $0.185-0.196$ and $0.251-0.282$ M$_\odot$ of \Ni{} respectively.

Therefore, we cannot assume that the amount of \Ni{} production depends only on the size and speed of the shock (which is larger for smaller initial velocities) but most likely also on the amount of energy in the initial system as well. This is in agreement with the result of case H0 in comparison to FF0, in addition to case FFS that produced more \Ni{} than the same velocity head-on case FF0. As we see in the next section, the amount of \Ni{} production also depends on the numerical resolution and numerical parameters of the system. We aim to study these effects in more detail in the future.

\subsection{Resolution and burning limiters study}
\label{sec:res}
\begin{table*}
    \centering
    \begin{tabular}{|c|c|c|c|c|c|c|c|}
    \hline
    \hline
     Simulation & Impact & $ V_\text{min}$ & limiters & $\Delta t_\text{min}$ & $\Delta t_\text{max}$ &  ${}^{28}$Si & ${}^{56}$Ni   \\
    Name  & parameter & ${10}^{19}\left[\text{cm}^3\right]$ & & ${10}^{-3}\left[\text{s}\right]$ & ${10}^{-3}\left[\text{s}\right]$ & $\left[\text{M}_{\odot}\right]$ & $\left[\text{M}_{\odot}\right]$ \\
    \hline
    \hline
     FF0 & 0 & $ 0.29 $ & no burning in shock & $0.45$ & $0.92$ & 0.487 & $0.267$ \\
     FF0-LRM & 0 & $160$ & no burning in shock & $1.83$ & $7.3$ & 0.514 & $0.162$ \\
     FF0-LRS & 0 & $160$ & no burning in shock & $1.46$ & $11.7$ & 0.484 & $0.21$ \\ 
     {FF0-BMB} & 0 & $5.03$ & no burning in shock & $ 0.915$ & $3.6 $ & 0.47 & $0.287$ \\ 
     {FF0-BM} & 0 & $4.8$ & no burning in shock & $ 0.915 $ &  $1.83 $ & 0.462 & $0.305$ \\ 
     {FF0-BMBT} & 0 & $ 1.2 $ & no burning in shock, temperature change & $ 0.457 $ & $ 1.8 $ & 0.309 & $ 0.699$ \\ 
     {FF0-MBM} & 0 & $47.1 $ & no burning in shock & $ 1.7 $ & $  6.9$ & 0.495 & $ 0.207$ \\ 
     FF0-LTS & 0 & $160$ & no burning in shock, nuclear time step limiter & $5.9 \cdot 10^{-4}$ & $5.9 \cdot 10^{-4}$ & 0.529 & $0.205$ \\
     FF0-LNS & 0 & $160$ & None & $2.44$ & $9.7$ & 0.597 & $0.07$ \\
    \hline
     {FFM-BMBT} & {$8.5\cdot {10}^8 $} & {$1$} & {no burning in shock, temperature change} & {$0.45$} & {$3.6$} & {0.307} & {0.677}\\ 
     {FFML-BMBT} & {$9.82 \cdot {10}^8$} & {$4.26 $} & {no burning in shock, temperature change} &{$0.45$} & {$3.6$} & {0.3} & {0.669} \\ 
     {FFL-BMBT} &  {$1.2\cdot {10}^9 $} & {$7.8$} & {no burning in shock, temperature change}& {$1.8$} & {$3.6$} & {0} & {0}\\
     \hline
     {H0-BMBT} & {$ 0 $} & {$2.3 $} & {no burning in shock, temperature change} & {$1.8$} & {$3.6$} & {0.227} & {0.944}\\ 
    \hline
    \hline
    \end{tabular}
    \caption{Initial parameters and results for the different simulations that initialized with a free fall velocity. $ V_\text{min}$ is the minimum cell volume along the simulation, indicating the resolution, limiters can be either no limiters, no burning in shock as described in \citet{2009SeitenzahlSpontaneousDetonationWD} and a time step limiter, as defined in \citet{2021PakmorZenatiHybridIgnition}. $\Delta t_\text{min}$ are the minimum time step of a cell in the simulation, $\Delta t_\text{max}$ is the maximum timestep of a cell in the simulation at the step of $\Delta t_\text{min}$,  ${}^{28}$Si and ${}^{56}$Ni are the final mass production of  ${}^{28}$Si and ${}^{56}$Ni. }
    \label{tab:simulations_parameters_resolution}
\end{table*}

We have also run several direct impact simulations using lower spatial resolution; different time limiters, and various types of burning limiters: no limiter, the limiter we used throughout the simulations as in \citet{2009SeitenzahlSpontaneousDetonationWD}, a nuclear timestep limiter where we reduce the hydro timestep so that the relative change of thermal energy in one timestep is small, and a limiter on the temperature change as in \cite{2013KushnirKatzHeadOnWDCollisionsInTriple}, where we reduce all nuclear reaction rates in a cell by a factor such that $\Delta \text{ln}T\le 0.1$ for a hydro timestep. The low-resolution runs (marked with "L" in their names in Tab. \ref{tab:simulations_parameters_resolution}), have a limit on the minimum volume to be $~1.6\cdot 10^{21} \text{cm}^3$ which is about 1/5 the size in the case of maximum resolution in \citet{2012HawleyHeadonFlashResolution}. We also ran 7 simulations with a factor of 8 bigger target mass, i.e. $m_\text{target}=1.6\cdot {10}^{27}$g (those with 'BM' in their names following the original simulation name in Tab. \ref{tab:simulations_parameters}), and the ones with the extra 'B', i.e.- 'BMB', have been initiated with the new target mass 1s before the impact (at 3.3s from the beginning of the simulation). In addition, those with 'T' at the end of their names (FF0-BMBT, FFM-BMBT, FFML-BMBT, FFL-BMBT, H0-BMBT) run with the same limit on the change in temperature as described in \cite{2013KushnirKatzHeadOnWDCollisionsInTriple}. Increasing the target mass of a cell induced lower spatial resolution. In addition, we ran a simulation with a factor of 16 bigger target mass (FF0-MBM), which we have run from the very beginning.  As can be seen in Tab. \ref{tab:simulations_parameters_resolution}, the various runs produce a range of \Ni{} masses with a factor of 4 in production between the lowest and highest \Ni{} production cases when running without the limiter on the temperature change. This is consistent with the results of \cite{2010RaskinNi56FromWDCollisions,2012HawleyHeadonFlashResolution}, showing that the results of \Ni{} production are resolution-dependent. When limiting the amount of temperature change allowed \citep{2013KushnirKatzHeadOnWDCollisionsInTriple}, we get a much higher \Ni{} production, consistent with \cite{2013KushnirKatzHeadOnWDCollisionsInTriple} and \cite{2010RaskinNi56FromWDCollisions}. To verify that there is no change in the detonation itself when running with this limiter, we ran the limiting cases with impact parameters close to the calculated threshold. The results in terms of the outcomes were similar, but the production of \Ni{} (in case of explosion) is more than a factor of 2 higher.
We present the amount of \Ni{} produced in past simulations in Tab. \ref{tab:past_simulations}. With 200k SPH particles, the simulation of similar masses by \cite{2010RaskinNi56FromWDCollisions} produced between $0.65 \text{M}_\odot$ to $0.84 \text{M}_\odot$ of \Ni. However, using the grid code \texttt{FLASH},\citep{2012HawleyHeadonFlashResolution} showed that the resolution affects the results, finding masses between $0.63 \text{M}_\odot$ for the lowest resolution and $0.39 \text{M}_\odot$ for their highest resolution runs. On the other hand, \cite{2013KushnirKatzHeadOnWDCollisionsInTriple} claims that the amount of \Ni{} should be in the range of $0.5-1 \text{M}_\odot$. We find that most comparable cases (with initial free-falling collisions) produce around $0.2 \text{M}_\odot$ of \Ni, with no clear correlation to the impact parameter (see Tab. \ref{tab:simulations_parameters}) nor to the different simulation configurations (Tab. \ref{tab:simulations_parameters_resolution}).
Given the demanding computational expenses of such 3D models, we could not further explore the range of impact parameters and velocities for these types of experiments.
The absence of a significant change in the amount of \Ni{} in H0-BMBT compared to H0 suggests that the low production of \Ni{} found in most of the free-falling cases where the temperature change limiter was not used, was likely due to an earlier detonation of the carbon, allowing more time for the collision shock front to compress the WDs, before the burning front runs over them. In contrast, in the hypervelocity head-on case, the collisional shock propagates faster than the burning front, reaching the edge of the objects and already compressing much of the WDs before the detonation, in both the case where a limiter was used and the case modeled without this limiter. 

\begin{figure*}
    \centering
    \begin{tabular}{l|l}
      \includegraphics[width=0.5\linewidth,clip]{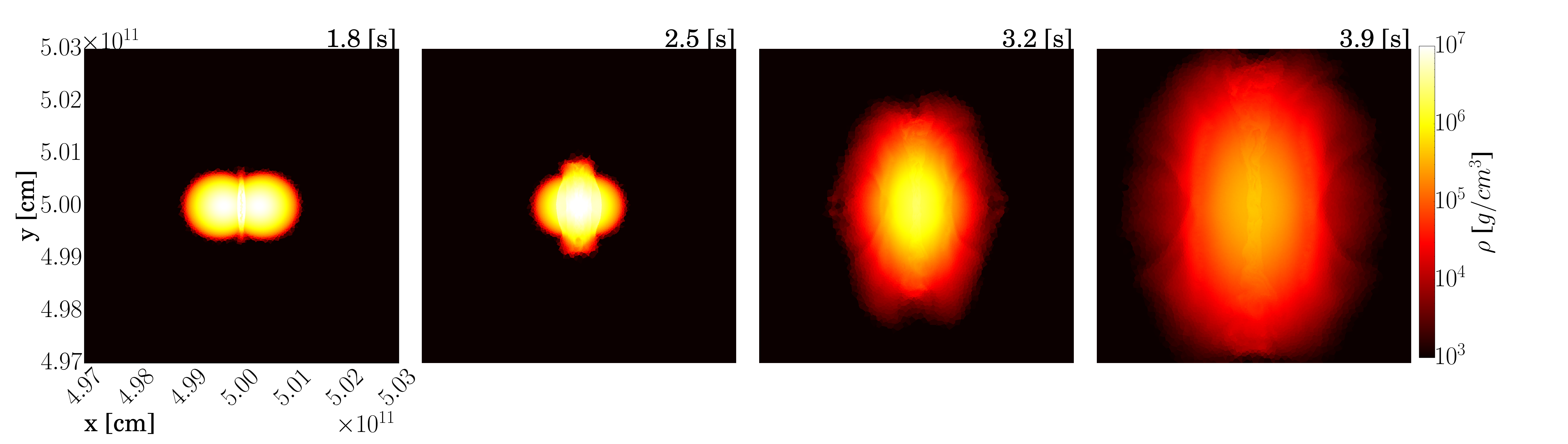} &
      \includegraphics[width=0.5\linewidth,clip]{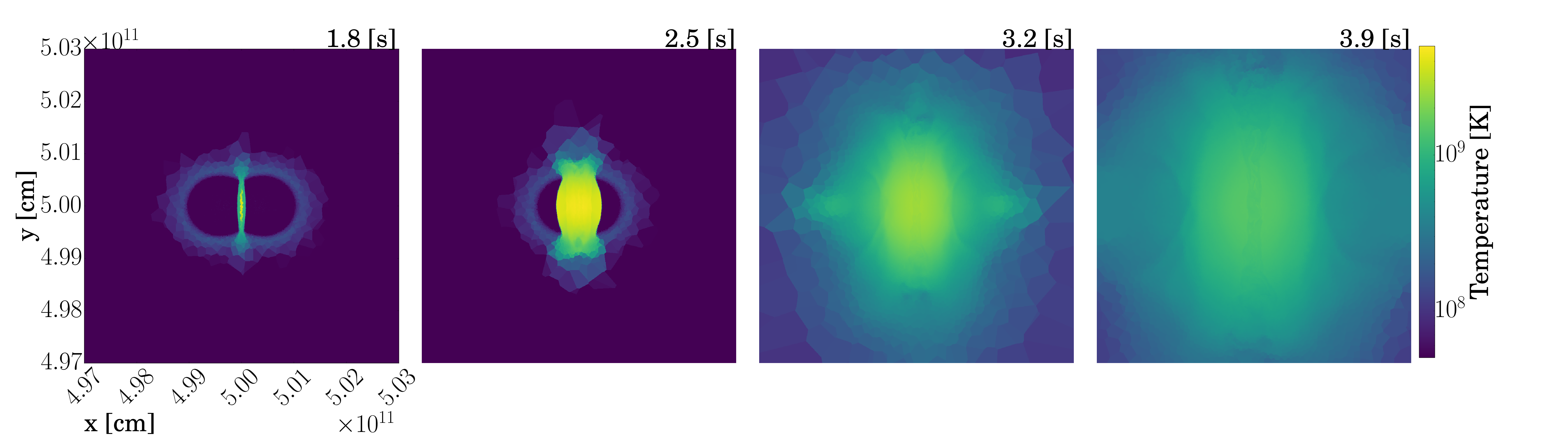}\\
    \includegraphics[width=0.5\linewidth,clip]{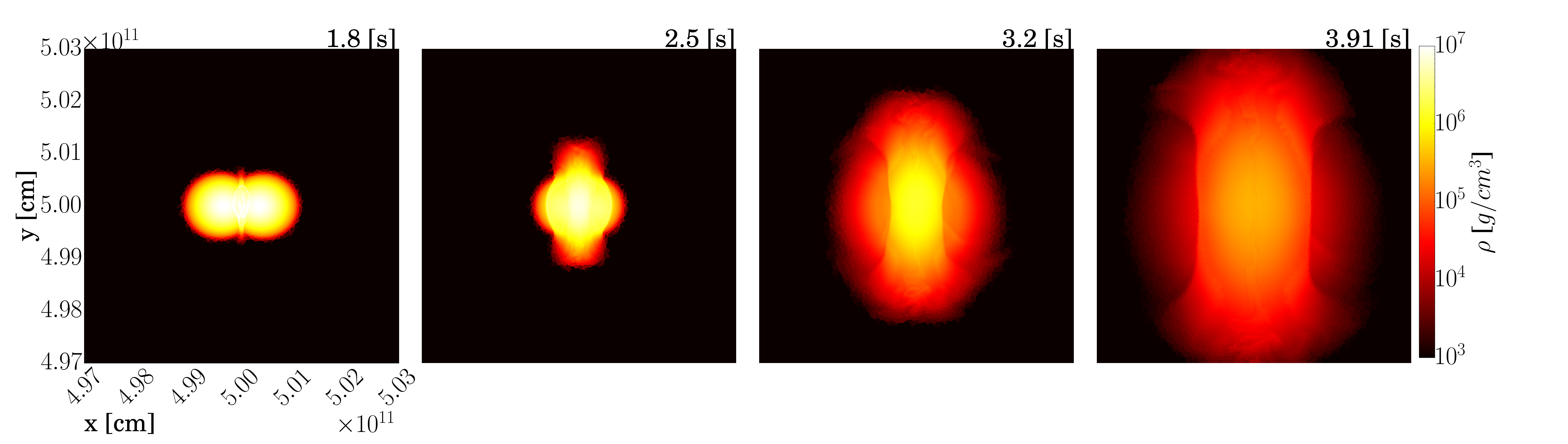} &
    \includegraphics[width=0.5\linewidth,clip]{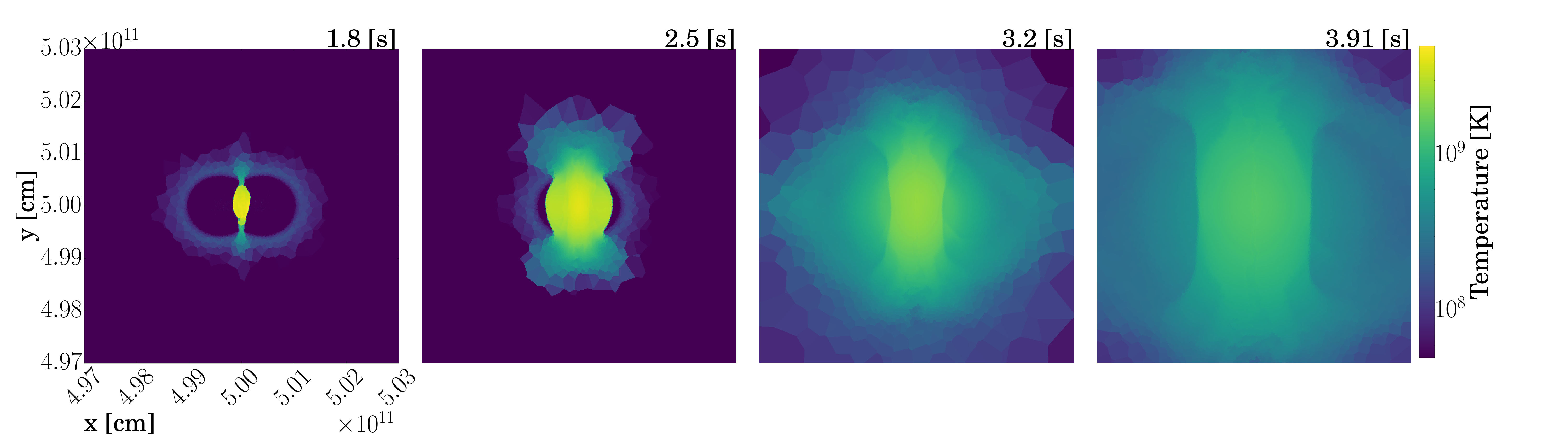}\\
    \hline \\
    \includegraphics[width=0.5\linewidth,clip]{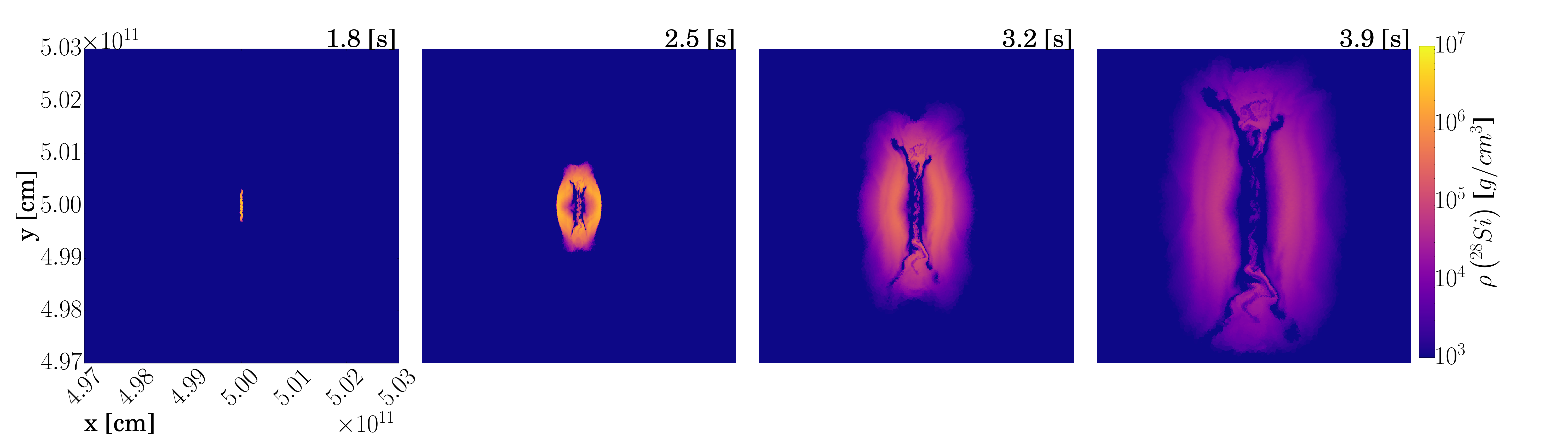} &
    \includegraphics[width=0.5\linewidth,clip]{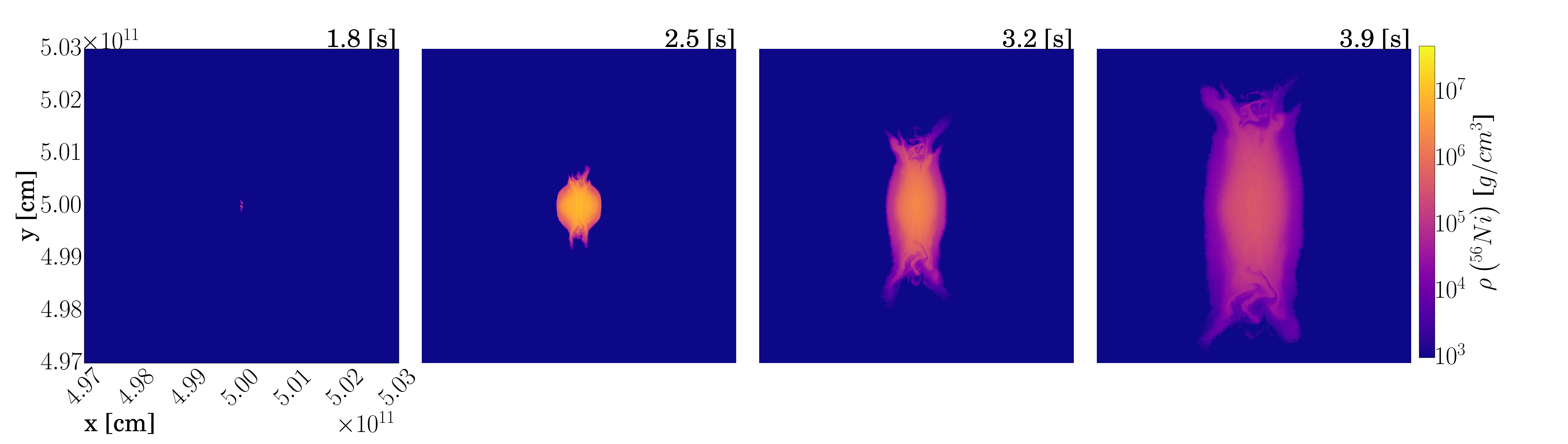}\\
    \includegraphics[width=0.5\linewidth,clip]{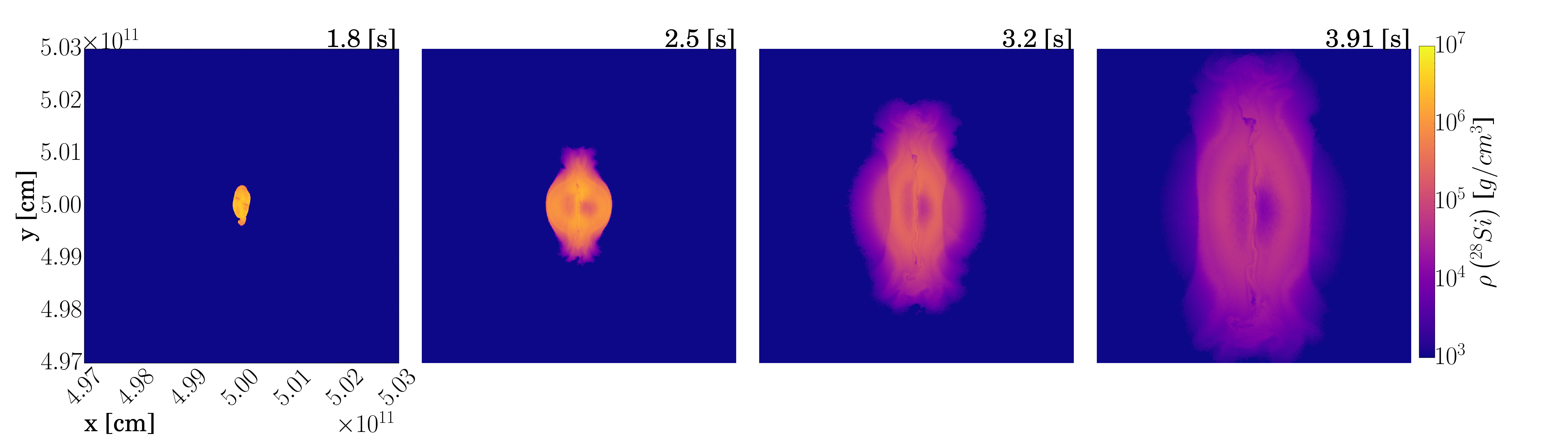} &
    \includegraphics[width=0.5\linewidth,clip]{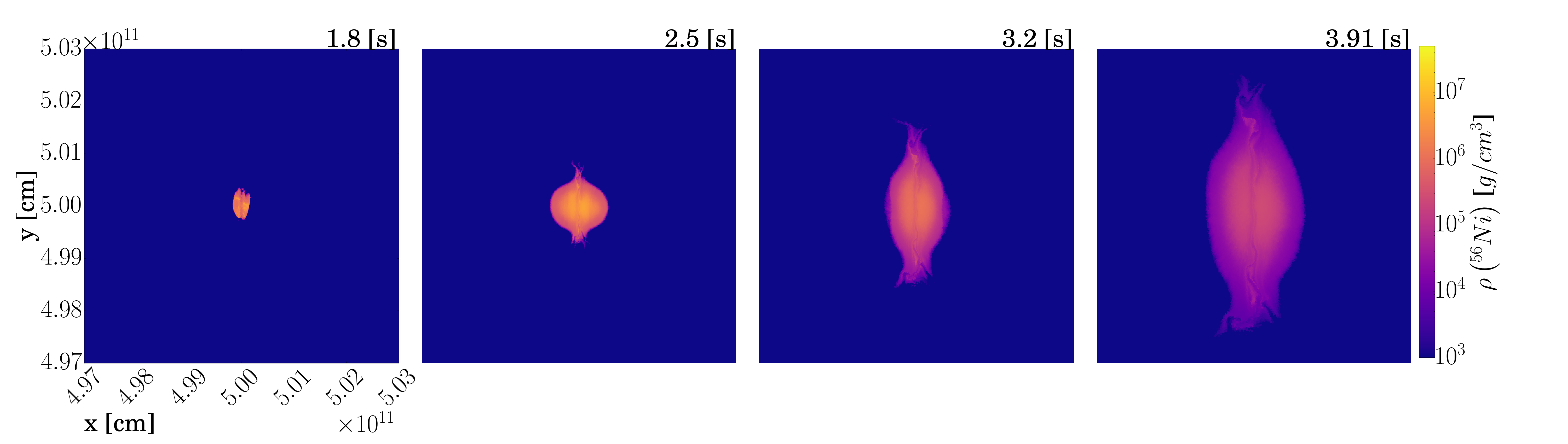}
    \end{tabular}
    
    \caption{The difference in shock propagations from the direct collision and the burning between the simulations running with (top panels in each quadrant, case FF0-BMBT) and without (bottom panels, case FF0-BMB) the limiter discussed in \cite{2013KushnirKatzHeadOnWDCollisionsInTriple}. }
    \label{fig:kushnir-temp}
\end{figure*}

\section{Discussion and Summary}
In this paper, we studied the direct physical collisions between two WDs at various impact parameters and velocities and their resulting explosions.

Our models suggest that beyond a critical impact parameter, the direct collisions of two WDs do not cause detonation. We have provided an analytic model to identify this threshold and found that it compares well with the results of our direct numerical simulations. This provides a simple way to account for the impact parameter in the estimates of SNe rates from WD collisions. Our results also show that the regime of impact parameters that lead to explosions is between 0.4 to 0.5 of the entire range of collision impact parameters. Therefore, existing rate estimates for SNe from WD collisions (which typically assume any physical collision gives rise to a detonation and produces a SN) are overestimated by about a factor 2-2.5. Overall, our results, as well as population synthesis studies, suggest that only a small fraction of Ia SNe are likely to arise from WD collisions.

Finally, similar to previous studies, we have also found that the amount of \Ni{} produced in such WD collisions depends on the resolution of the simulations and the specific burning limiters used \citep{2010RaskinNi56FromWDCollisions, 2012HawleyHeadonFlashResolution, 2013KushnirKatzHeadOnWDCollisionsInTriple}.
While testing the \Ni{} production with different limiters, we find that when the burning shock is delayed in comparison to the collision shock, more of the WDs can reach the critical properties for detonation, and therefore eventually more \Ni{} is produced. This is possibly the reason why there is much more \Ni{} produced in the H0 case, when the collision shock is much faster than the burning front, unlike the cases with $b\ne 0$ and those with lower velocities. The collision shock front compresses material and raises the temperature. Once the detonation occurs, the burning front propagates throughout the WD. The burning efficiency now depends on the density of the material, and for full burning up to \Ni{}, one requires burning material at densities typically about $10^7 \gcms$. The innermost region of the WD already has such density, but other parts only achieve this after the collision shock front compresses them. In general, the amount of \Ni{} produced in these events will affect the light curve and spectra, and is, therefore, important to further investigate if one wants to quantify the actual abundances of SNe Ia.

\begin{table*}
    \centering
    \begin{tabular}{|c|c|c|c|c|c|c|c|c|c|}
    \hline
    \hline
     Paper & Masses &b& code & type of hydro code & $V_\text{min}$ & $\rhoint$ & $u_\text{relative}$ & ${}^{56}$Ni   \\
     & $\left[\text{M}_\odot\right]$ & $\left[R_{WD}\right]$  & & & ${10}^{19}\left[\text{cm}^3\right]$ & ${10}^{5}\left[\text{g} / {\text{cm}}^3\right]$ &${10}^{8}\left[\text{cm} / \text{s}\right]$ & $\left[\text{M}_{\odot}\right]$ \\
    \hline
    \hline
     \cite{2009RosswogWDCollision} & $0.6 + 0.6$ & $0$ &\texttt{FLASH} & 3D AMR & $12$ &$34$& $4.32$& $0.16$ \\
     \cite{2009RosswogWDCollision} & $0.6 + 0.6$ & $0$ & own code& 3D SPH &SPH&$~34$& $4.32$& $0.32$\\
     \cite{2009RaskinWDCollisionSimulation} & $0.6 + 0.6$ & $0$ & \texttt{SNSPH} & 3D SPH &SPH&$34$& $4.32$& $0.4$ \\
     \cite{2010RaskinNi56FromWDCollisions} &  $0.64 + 0.64$ & $0$ & \texttt{SNSPH} & 3D SPH &SPH& $40$ & $~4.52$ &  $0.51$ \\
     \cite{2010RaskinNi56FromWDCollisions} & $0.81 + 0.81$ & $0$ & \texttt{SNSPH} & 3D SPH &SPH& $100$ & $~5.59$ & $0.84$\\
     \cite{2012HawleyHeadonFlashResolution} &  $0.64+0.64$ & $0$ & \texttt{FLASH} & 3D AMR  & $219$ &$40$& $4.52$ & $0.32$ \\
     \cite{2012HawleyHeadonFlashResolution} &  $0.81+0.81$ & $0$ & \texttt{FLASH} & 3D AMR  & $126$ & $100$& $5.59$ & $0.39$\\
     \cite{2013GarciaHeadOnSimulations} & $0.7 + 0.7$ & $0$ & \texttt{AXISSPH} & 3D SPH &SPH&$60$& $4.85$& $0.42$ \\
    \cite{2013KushnirKatzHeadOnWDCollisionsInTriple} & $0.64+0.64$  & $0$ & \texttt{FLASH / } & 2D AMR /  & $~3\cdot {10}^{-3}$& $~40$ & $~4.52$ & $0.41$\\
      &  &  & \texttt{VOLCAN2D} & 2D ALE &  &  &  & \\
    \cite{2013KushnirKatzHeadOnWDCollisionsInTriple} &  $0.8+0.8$  & $0$ & \texttt{FLASH / } & 2D AMR /  & $~3\cdot {10}^{-3}$& $100$& $~5.5$ & $0.74$ \\
      &  &  & \texttt{VOLCAN2D} & 2D ALE &  &  &  & \\
     \cite{2016PapishPeretsHeadOnCollision} & $0.6 + 0.6$ & $0$ &\texttt{FLASH} & 2D AMR & $0.1$ & $34$ & $4.32$& $0.31$\\
     \cite{2016PapishPeretsHeadOnCollision} & $0.7 + 0.7$ & $0$ &\texttt{FLASH} & 2D AMR & $0.1$ & $59$ & $4.85$& $0.52$ \\
     \hline
     \cite{2009RaskinWDCollisionSimulation}&  $0.6 + 0.6$ & $2$ & \texttt{SNSPH} & 3D SPH &SPH& $0.05 - 10$ &$4.32$& - \\
     \cite{2010RaskinNi56FromWDCollisions} &  $0.64 + 0.64$ & $2$ & \texttt{SNSPH} & 3D SPH &SPH& $0.05 - 10$ & $~4.52$ &  -\\
     \cite{2010RaskinNi56FromWDCollisions} & $0.81 + 0.81$ & $2$ & \texttt{SNSPH} & 3D SPH &SPH& $0.7-10$ & $ 5.59$ & $0.65$\\
    \hline
    \hline
    \end{tabular}
    \caption{Initial parameters and results for the different previous simulations. We present b, the impact parameter, $\rhoint$, and $u_\text{relative}$ upon collision. $ V_\text{min}$ is the minimum cell volume along the simulation, indicating the resolution; we present this value only for the simulations that used mesh codes. ${}^{56}$Ni is the final mass production of ${}^{56}$Ni, which is not presented in case of no explosion. Values that were not directly specified in the papers were calculated assuming a free-fall velocity and the density profiles presented there or, if not given- from our Fig. \ref{fig:density-r}. The last 3 values of $\rhoint$ are not directly given in the papers and, therefore, are not clear. }
    \label{tab:past_simulations}
\end{table*}

Since the radius at which the critical density is located is about the same for every CO WD that we tested with \texttt{MESA} (Fig. \ref{fig:density-r}), and is about $d_\text{crit}\approx 5.1 \cdot 10^8 \cm$, we can deduce that collisions of more massive WDs will produce more explosions with more \Ni{} production. However, even for a grazing collision, the impact parameter should be small, as they are more compact.

From Fig. \ref{fig:ejecta}, we can infer that the resulting ejecta of non-zero impact-parameter collisions is not symmetric and might show unique signatures in the spectra,  polarization, and supernova remnant structure. Follow-up studies of these could provide important information in this regard. It is, however, beyond the scope of the current study.

\section*{Acknowledgements}
HBP acknowledges support for this project from the European Union's Horizon 2020 research and innovation program under grant agreement No 865932-ERC-SNeX. HG acknowledges support for the project from the Council for Higher Education of Israel.  
\section*{Data Availability}
Requests for data not in the Zenodo repository is available upon reasonable request to the corresponding authors.

\bibliography{references}{}
\bibliographystyle{aasjournal}



\end{document}